\documentclass[a4paper,12pt]{article}
\usepackage{amsmath,amssymb,amsfonts,epsfig,graphicx}
\usepackage{colordvi}
	\fboxrule 0.2mm 	\newcounter{comentario} 	

\title{Phases of the electronic two-level model under the rotating wave approximation}

\author{ M.T. Thomaz$^{1}\footnote{Corresponding author: mtt@if.uff.br}$,
 A.C. Aguiar Pinto$^{2}$ and M. Moutinho$^{2}$   \vspace{0.25cm} \\ 
\small\it $^{1}$Instituto de F\'{\i}sica, Universidade Federal Fluminense,\\ 
\small\it Av. Gal. Milton Tavares de Souza s/n$^{\textit o}$, 
CEP 24210-346, Niter\'oi-RJ, Brazil 
\vspace{0.25cm} \\
\small\it $^{2}$
Curso de F\'{\i}sica, Universidade Estadual de Mato Grosso do Sul,\\
\small\it Caixa Postal: 351, Cidade Universit\'aria de Dourados, CEP 79804-970, 
Dourados-MS,  Brazil.
}

\begin{document}

\maketitle

\begin{abstract}

We present the time evolution of the electronic two-level model in the
rotating wave approximation (RWA). We calculate the Aharonov-Anandan
phase of the vector states with cyclic evolution either within the
period of an external monochromatic electric field or the period
corresponding to Rabi's frequency. The Aharonov-Anandan phase is
shown to be dependent on the initial vector state, unless the system
evolves in the adiabatic regime; in the latter case, the
Aharonov-Anandan phase recovers Berry's phase. Our results are also
discussed in the quasi-resonant regime.

\end{abstract}

\vfill



\newpage


\section{Introduction} \label{S1}

A quantum system coupled to a classical monochromatic laser
in the quasi-resonant regime can be approximated  by a $N$-level
model, $N$ being finite, in the rotating wave
approximation (RWA)\cite{livro_optica}.

The RWA of the two-level model
has been used as a candidate for modeling geometric phase 
gates between ion-qubits in the presence of an external electric 
or magnetic field\cite{nielsen}. These two-level
models can be realized through the interaction of an external magnetic
field with a spin-$1/2$\cite{rabi1936,rabi1937},
or matter coupled to an external electric field, in the
dipole approximation\cite{livro_optica}.

Recently we have studied the adiabatic evolution of a two-level
model\cite{BJP09} ({\it i}) under a (real) classical monochromatic
electric field and ({\it ii}) when only the contribution of the
positive frequency of the electric field is taken into
account (its RWA). In the RWA of the two-level model we 
recover the geometric phases acquired by the instantaneous 
eigenstates of energy already known in
the literature\cite{nao-hermitiana}. On the other hand, we 
verified the absence of geometric phase in the adiabatic 
evolution of the instantaneous eigenstates of the two-level 
model coupled to a real monochromatic electric field.

Let $\omega$ be the angular frequency of the monochromatic
external electric field and $\Delta \varepsilon$ the
energy difference of the two-level model. The
matter-field coupling can only be approximated by two-level 
model in the quasi-resonant regime\cite{livro_optica}, in which
$\omega \approx \Delta \varepsilon$
(in natural units, $c= \hbar =1$). Certainly this condition
does not fulfill  the adiabatic condition\cite{BJP09} 
$\omega \ll \Delta \varepsilon$. In 2002, Xin {\it et al}.\cite{xin}
proposed a scheme to perform quantum geometric computation in
 non-adiabatic regime with trapped ions.

Recently Imai {\it et al}.\cite{imai1} and Imai and
Morinaga\cite{imai2} measured the geometric phases in the two-level
model coupled to a monochromatic electric field. In Refs.
\cite{imai1,imai2} it is mentioned that "{\it A geometric phase solely
depends on the amount of the solid angle enclosed by the evolution
path, not on the details ..., or the initial and final states of the
evolution.}". This statement is certainly true when the quantum model
is under adiabatic evolution\cite{berry}, but it is not obvious
that this comment is valid in the RWA of the two-level model coupled
to a monochromatic electric field in the quasi-resonant regime
($\omega = \Delta \varepsilon + \delta$, and $\frac{\delta}{\Delta
\varepsilon} \ll 1$).

The present work aims at studying the dynamics
the two-level model under the RWA and coupled
to a linearly polarized
monochromatic electric field, in the electric dipole
approximation\cite{livro_optica}, for any value of the
ratio $\frac{\omega}{\Delta \varepsilon}$.
We want to determine the initial conditions
under which a vector state acquires an
Aharonov-Anandan phase\cite{aharonov} within different
periods.  Our model is simpler than the one studied in 
Refs. \cite{imai1,imai2}, where different laser beams  
are applied to matter in order that the quantum system
acquire the geometric phases.

In section \ref{S2} we present the hamiltonian of
model under study. Its dynamics is derived
by mapping it onto a spin-$1/2$ model coupled to an effective
magnetic field. The study of global phases in the non-adiabatic 
evolution of  the spin-1/2 model in the presence of a 
rotating magnetic field was carried out previously in 
Refs.\cite{bulgac,ni,zhu}. In section \ref{S3} we study the cyclic
evolution of the vector states under the action of the
electric two-level model. The Aharonov-Anandan phase
acquired by the states are considered for state vector with 
periodicity within that of the external electric field
or that of Rabi's frequency. Our results can be used to extend 
easily  the discussions made in Refs.\cite{bulgac,ni,zhu} on 
these phases in the spin-1/2 model  under the action of a 
precessing magnetic field. A summary of our results
is presented in section \ref{S4}. \ref{Appendix_A} presents 
the average energy in any vector state.


\section{ The dynamics of the RWA of the two-level model } \label{S2}

We begin with the hamiltonian ${\bf H}_0$ of the quantum two-level model,
without interaction\cite{cohen}, 

\begin{eqnarray} \label {2.1}
{\bf H}_0 = \varepsilon_1 |1\rangle \langle 1| 
           + \varepsilon_2 |2\rangle \langle 2| , 
\end{eqnarray}

\noindent where ${\bf H}_0 |i \rangle = \varepsilon_i |i \rangle$,
$i= 1,2$. We take $\varepsilon_1 < \varepsilon_2$, which
prevents (\ref{2.1}) from describing a degenerate
two-level system. We use the natural units: $c= \hbar=1$.

The two-level interaction hamiltonian of the matter and the 
electric field, in the electric dipole approximation, 
${\bf H}_e (t)$,  represented in the basis of the eigenstates 
of ${\bf H}_0$, is \cite{lamb}

\begin{eqnarray}  \label{2.2}
  {\bf H}_e (t) = \left(
     \begin{array}{c c}
     \varepsilon_1  &  \vec{d}_{12} \cdot \vec{E} (t) \\
      (\vec{d}_{12} \cdot \vec{E} (t))^{*} & \varepsilon_2
     \end{array}
  \right) ,
\end{eqnarray}

\noindent  in which

\begin{eqnarray}  \label{2.3}
  \vec{d}_{12} \equiv -e  \, \langle 1| \vec{\bf x} | 2 \rangle ,
\end{eqnarray}

\noindent and $\vec{\bf x}$ is the position operator of the electron.
We assume that the eletronic states $|i \rangle$, $i= 1,2$,
are spherically symmetric, and hence 
do not have a permanent electric dipole ($\vec{d}_{ii} =0$, $i=1,2$).
Throughout our calculations, we have chosen the basis of 
eigenstates of ${\bf H}_0$, such that  the components of  
$\vec{d}_{12}$ are  real; we have taken $e>0$, so that the 
electronic charge is $-e$. 

In the RWA, the main contribution to the quantum phenomenon comes from the 
positive angular frequency of the monochromatic electric field. In this
approximation the classical electric field in hamiltonian  (\ref{2.2}) 
becomes

\begin{eqnarray} \label{2.4}
\vec{E} (t) \approx \vec{E}_0 \; e^{-i (\omega t - \phi_0)} ,
\end{eqnarray}

\noindent where $\vec{E}_0 \in \mathbb{R}^3$, $\omega > 0$  and the 
phase $\phi_0$ is chosen such  that $\vec{d}_{12}\cdot\vec{E}_0 \ge 0$.
For the complex electric field (\ref{2.4}), the hamiltonian  (\ref{2.2})
depends on the two dimensional  variable
($\vec{d}_{12}\cdot\vec{E} (t)$).

The dynamics of the spin-$1/2$ coupled to a magnetic field that 
precess around a fixed direction has been studied previously
for any dynamical regime\cite{ajp2000,gonzalo}.

By making a shift in hamiltonian (\ref{2.2}), it can be written 
as the hamiltonian of a spin-$1/2$ coupled to an effective 
magnetic field, that is,

\begin{eqnarray}  \label{2.5}
{\bf H}_e (t) = \frac{(\varepsilon_1 + \varepsilon_2) \mathbf{1}}{2}
+ \frac{\mu}{2} \vec{B}_{eff} (t) \cdot \vec{\sigma},
\end{eqnarray}

\noindent where $\mathbf{1}$ is the identity  operator of dimension 
$2\times 2$, $\vec{\sigma}$ are the Pauli matrices, $\mu = \mu_B g$, 
with $g$ being the Land\'e's factor and $\mu_B$  the Bohr magneton.
The components of the effective magnetic field on the r.h.s. of 
eq.(\ref{2.5}) are,

\begin{subequations}

\begin{eqnarray}  \label{2.6}
B_x^{eff} (t) &=& \frac{2D_0}{\mu} \cos( \omega t - \phi_0),  \label{2.6a}\\
B_y^{eff} (t) &=&  \frac{2D_0}{\mu} \sin( \omega t - \phi_0),  \label{2.6b}\\
B_z^{eff} (t) &=& - \frac{\Delta \varepsilon}{\mu},  \label{2.6c}
\end{eqnarray}

\end{subequations}

\noindent where

\begin{subequations}

\begin{eqnarray}  \label{2.7a}
D_0 \equiv \vec{d}_{12} \cdot \vec{E}_0 \ge 0 
\end{eqnarray}

\noindent and 

\begin{eqnarray} \label{2.7.b}
\Delta \varepsilon &\equiv& \varepsilon_2 - \varepsilon_1 >0 .  
\end{eqnarray}

\end{subequations}

The eigenvectors of hamiltonian ${\bf H}_0$  ($|i\rangle$, $i=1, 2$) 
are also eigenvectors of the operator $\sigma_z$, that is,
$\sigma_z |1\rangle = |1\rangle$ and 
$\sigma_z |2\rangle = - |2\rangle$.

The components of the effective magnetic field $\vec{B}_{eff} (t)$ 
can also be written in terms of polar coordinates, 
$\vec{B}_{eff} (t) = (B \sin(\theta) \cos(\omega t^{\prime}), 
B \sin(\theta) \sin(\omega t^{\prime}), B \cos(\theta))$, where
$t^{\prime} \equiv t - \frac{\phi_0}{\omega}$ and 

\begin{subequations}

\begin{eqnarray} \label{2.8}
\cos(\theta) &=& - \frac{1}{2} \; 
\frac{\Delta \varepsilon}{\sqrt{D_0^2 + (\frac{\Delta \varepsilon}{2})^2}} ,
                             \label{2.8a}  \\
   \nonumber \\
\sin(\theta) &=& \frac{D_0}{\sqrt{D_0^2 + (\frac{\Delta \varepsilon}{2})^2}} .
                             \label{2.8b}
\end{eqnarray}

\end{subequations}

The norm $B$ of the effective magnetic field is

\begin{eqnarray}  \label{2.9}
B = \frac{2}{\mu} \, \sqrt{D_0^2 + (\frac{\Delta \varepsilon}{2})^2} .
\end{eqnarray}

The effective magnetic field on the r.h.s. of hamiltonian (\ref{2.5})
has  constant norm and constant $z$-component. It precesses 
around the $z$ direction with angular velocity $\omega$. From 
expressions (\ref{2.8a}) and (\ref{2.8b}), we conclude that 
$\theta \in [\frac{\pi}{2}, \pi]$. 

The eigenvectors and the eigenvalues of the second term on the
r.h.s. of hamiltonian (\ref{2.5}) were calculated in 
Refs.\cite{bulgac,ni,ajp2000,gonzalo}. The eigenvectors of the hamiltonian of 
a spin-$1/2$  coupled to a precessing magnetic field calculated in 
these two references differ by global phases. From now on, 
we follow  Ref.\cite{ajp2000}.

The eigenvalue equation of ${\bf H}_e (t)$ is

\begin{eqnarray}  \label{2.10}
{\bf H}_e (t) \, | \phi_i; i \rangle = E_i \, | \phi_i; i \rangle, \hspace{0.5cm}
i= 1, 2 ,
\end{eqnarray}

\noindent where

\begin{subequations}

\begin{eqnarray}  \label{2.11a}
E_i = \frac{( \varepsilon_1 + \varepsilon_2)}{2} + \tilde{E}_i,
\hspace{0.5cm} i = 1, 2,
\end{eqnarray}

\noindent with 

\begin{eqnarray}   \label{2.11b}
\tilde{E}_1 = - \,  \sqrt{D_0^2 + (\frac{\Delta \varepsilon}{2})^2}
\hspace{0.5cm} \mbox {and} \hspace{0.5cm}
\tilde{E}_2 =  \sqrt{D_0^2 + (\frac{\Delta \varepsilon}{2})^2} .
\end{eqnarray}

\end{subequations}

The eigenvectors of ${\bf H}_e (t)$ associated to $E_1 (t)$
and $E_2 (t)$ are respectively:

\begin{subequations}

\begin{eqnarray}
|\phi_1; t\rangle &=& - \sin(\frac{\theta}{2}) | 1 \rangle 
   + \cos(\frac{\theta}{2})\; e^{i (\omega t - \phi_0)} | 2 \rangle  ,
                                           \label{2.12a} \\
 && \nonumber  \\
|\phi_2; t\rangle &=&  \cos(\frac{\theta}{2}) | 1 \rangle 
   + \sin(\frac{\theta}{2})\; e^{i (\omega t - \phi_0)} | 2 \rangle .
                                           \label{2.12b}
\end{eqnarray}

\end{subequations}

From eqs.(\ref{2.8a}) and (\ref{2.8b}) we obtain

\begin{subequations}

\begin{eqnarray}
\cos (\frac{\theta}{2}) &= &\frac{1}{\sqrt{2}} \;
\left[ 1 - \frac{\left( \frac{\Delta \varepsilon}{2}\right)}{\sqrt{D_0^2 
         + (\frac{\Delta \varepsilon}{2})^2}} \right]^{1/2} ,
                       \label{2.13a} \\
 &&  \nonumber \\
\sin (\frac{\theta}{2}) &= &\frac{1}{\sqrt{2}} \;
\left[ 1 + \frac{\left( \frac{\Delta \varepsilon}{2}\right)}{\sqrt{D_0^2 
         + (\frac{\Delta \varepsilon}{2})^2}} \right]^{1/2} .
                       \label{2.13b}
\end{eqnarray}

\end{subequations}

From the previous expressions of the eigenvalues and of the eigenvectors
of ${\bf H}_e (t)$, we verify that the ratio 
$(\frac{2 D_0}{\Delta \varepsilon})$ characterizes  the coupling strength  
between  the  matter and the external electric field: {\it i}) 
$\frac{2 D_0}{\Delta \varepsilon} \ll 1$, means a weak coupling;
{\it ii}) $\frac{2 D_0}{\Delta \varepsilon} \gg 1$, means to a strong
coupling. We remind that the strength of the coupling depends not only
upon the norm of the electric field but also on its relative direction 
to the vector $\vec{d}_{12}$ (see eq.(\ref{2.7a})).

\vspace{0.5cm}

Let us consider the initial vector state $|\psi (0) \rangle$,

\begin{eqnarray}
|\psi (0) \rangle = c_1 (0) | \phi_1; 0\rangle + c_2 (0) | \phi_2; 0\rangle,
       \label{2.14}
\end{eqnarray}

\noindent where $c_1 (0)$ and $c_2 (0) \in \mathbb{C}$.
The states $| \phi_i; 0\rangle$, $i = 1, 2$, are equal to the states
eqs.(\ref{2.12a}) and (\ref{2.12b}) at $t=0$. These coefficients
satisfy the normalization condition: $|c_1 (0)|^2 + |c_2(0)|^2 =1$. 

The time evolution of the vector $|\psi (t) \rangle$ is 

\begin{eqnarray}    \label{2.15}
{\bf H}_e (t) \, |\psi (t) \rangle = i \, \frac{d |\psi (t) \rangle}{dt},
\end{eqnarray}

\noindent and it is subjected to the initial condition (\ref{2.14}).

Using the instantaneous eigenstates of energy (\ref{2.12a}) and
(\ref{2.12b}) to decompose the vector $| \psi (t) \rangle$,

\begin{eqnarray}  \label{2.16}
| \psi (t) \rangle = \sum_{j=1}^{2} \; c_j (t) \, e^{-i E_j t} \; | \phi_j; t\rangle,
\end{eqnarray}

\noindent and substituting  eq.(\ref{2.16}) in the Schr\"odinger 
equation (\ref{2.15}), we obtain the equations satisfied by the 
coefficients $c_1 (t)$ and $c_2 (t)$. These equations are identical
to eqs. (12) and (13) of Ref.\cite{ajp2000} if in the latter  equations 
we make the following replacements: $\omega_1 = \frac{E_2 - E_1}{2}$ 
and $\omega_0 =  \omega$.

To simplest expressions of the coefficients, we 
include the dynamical phases in them and define 

\begin{eqnarray}  \label{2.17}
\tilde{c}_j (t) \equiv c_j (t) \; e^{-i E_j t}, 
    \hspace{0.3cm} j= 1, 2 .
\end{eqnarray}

\noindent The expressions of the coefficients $\tilde{c}_j (t)$, $ j=1,2$, at any 
instant $t$, are

\begin{subequations}

\begin{eqnarray}
\tilde{c}_1 (t) &=& e^{- \frac{i}{2} (\varepsilon_1 + \varepsilon_2 + \omega)t} \;
\left\{ c_1(0) \cos(\Gamma t)  
+ i \frac{\sin(\Gamma t)}{\Gamma} \left[( \tilde{E}_2 
              - \frac{\omega}{2} \cos(\theta)) \, c_1(0)\right. \right.  \nonumber \\
&& \hspace{1.5cm} \left. \left. 
- \frac{\omega}{2} \sin(\theta) \, c_2 (0) \right] \right\}  
                                \label{2.18a}  \\
 &&  \nonumber \\
\tilde{c}_2 (t) &=& e^{- \frac{i}{2} (\varepsilon_1 + \varepsilon_2 + \omega)t} \;
\left\{ c_2(0) \cos(\Gamma t)  
- i \frac{\sin(\Gamma t)}{\Gamma} \left[( \tilde{E}_2 
              - \frac{\omega}{2} \cos(\theta)) \, c_2(0)\right. \right.  \nonumber \\
&& \hspace{1.5cm} \left. \left. 
+ \frac{\omega}{2} \sin(\theta) \, c_1 (0) \right] \right\} ,
                                \label{2.18b}  
\end{eqnarray}

\end{subequations}

\noindent where $\Gamma$ is the Rabi's frequency and

\begin{subequations}

\begin{eqnarray}  
\Gamma &= & \frac{1}{2} \; \sqrt{(\mu B - \omega \cos(\theta))^2 + \omega^2 \sin^{2} (\theta)}    
         \label{2.19a} \\
 &=& \sqrt{D_0^2 + \frac{1}{4} \, (\Delta \varepsilon + \omega)^2} .
         \label{2.19} 
\end{eqnarray}

\end{subequations}

\vspace{0.3cm}

The expressions (\ref{2.18a}) and (\ref{2.18b}) are valid for any value
of the ratio $(\frac{\omega}{\Delta \varepsilon})$, including the 
quasi-resonant regime.


\vspace{0.5cm}

\section{ The Aharonov-Anandan phase } \label{S3}

In 1987 Aharonov and Anandan extended the concept of 
geometric phase to any time evolution of a quantum
system\cite{aharonov}. Suppose the vector state 
$|\psi (t)\rangle$ evolves under the action of 
an hamiltonian ${\bf H} (t)$. The cyclic evolution
of the quantum system is related to its vector state
during the period  $\tau$, and

\begin{eqnarray}   \label{3.1}
|\psi (\tau)\rangle = e^{i \phi} \, |\psi (0)\rangle  .
\end{eqnarray}

In Ref.\cite{aharonov}, the authors define a shifted
vector state $|\tilde{\psi} (t) \rangle$,

\begin{eqnarray} \label{3.2}
|\tilde{\psi} (t) \rangle \equiv  e^{- i f(t)} 
\; | \psi (t)\rangle,
\end{eqnarray}

\noindent where the function $f(t)$ satisfies the 
condition: $f(\tau) - f(0) = \phi$.

It is shown in this reference that

\begin{subequations}

\begin{eqnarray} \label{3.3a}
\phi = \beta - \int_0^{\tau} \, 
\langle \psi (t)| {\bf H} (t) | \psi (t) \rangle dt .
\end{eqnarray}

\noindent The expression of the Aharonov-Anandan phase,
the  $\beta$ phase,  is

\begin{eqnarray}  \label{3.3b}
\beta = \int_0^{\tau} \,
\langle \tilde{\psi} (t)| i \frac{d}{dt} \left( |\tilde{\psi} (t)\rangle \right) ,
\end{eqnarray}

\end{subequations}

\noindent that is formally identical to the phase acquired by 
the instantaneous eigenstates of energy when they evolve 
adiabatically\cite{berry}. We point out that differently from
the Berry's phase, the vector state in eq.(\ref{3.1}) is any
state that satisfies this cyclic condition.

Any initial vector state can be written as eq.(\ref{2.14}) and 
at any instant $t\ge 0$ it becomes

\begin{eqnarray} \label{3.4}
| \psi (t) \rangle = \sum_{j=1}^2 \; \tilde{c}_j (t)
\, | \phi_j; t \rangle .
\end{eqnarray}

\noindent The expressions of the coefficients 
$\tilde{c}_1 (t)$ and $\tilde{c}_2 (t)$ are given by
eqs. (\ref{2.18a}) and (\ref{2.18b}) respectively, and
$|\phi_j; t\rangle$, $j= 1, 2$, are the instantaneous
eigenstates of ${\bf H}_e (t)$ (see eqs.(\ref{2.12a})
and (\ref{2.12b})). The coefficients $\tilde{c}_j (t)$,
$j= 1, 2$, satisfy the normalization condition: 
$|\tilde{c}_1 (t)|^2 + |\tilde{c}_2 (t)|^2 =1$.

Along an adiabatic evolution, the coefficient 
$\tilde{c}_j (t)$, $j= 1, 2$, differs from its respective
initial coefficient $c_j (0)$, $j= 1, 2$, by at most a 
phase. That is not the general case of eqs. (\ref{2.18a})
and (\ref{2.18b}), where for arbitrary instant $t$ 
we have: $|\tilde{c}_j (t)| \not= |c_j (0)|$, $j= 1, 2$.

\vspace{0.5cm}

Let us call $T$ the period of the external electric field 
($\omega T = 2 \pi$). The hamiltonian (\ref{2.2}) has
also period $T$, ${\bf H}_e (T) = {\bf H}_e (0)$, as well
as its instantaneous eigenvectors.

We redefine the coefficients (\ref{2.18a}) and 
(\ref{2.18b}) as

\begin{subequations}

\begin{eqnarray}  
\tilde{c}_1 (t) &\equiv & e^{- \frac{i}{2} (\varepsilon_1 + \varepsilon_2 + \omega)t}
\; g_1 (t),   \label{3.5a}    \\
  &&   \nonumber \\
\tilde{c}_2 (t) &\equiv & e^{- \frac{i}{2} (\varepsilon_1 + \varepsilon_2 + \omega)t}
\; g_2 (t) .   \label{3.5b} 
\end{eqnarray}
 
\end{subequations}

We call $T_{\Gamma}$ the period associated to the Rabi's
frequency $\Gamma$ ( $\Gamma T_{\Gamma} = 2 \pi$). From
the definitions of $g_j (t)$, $j= 1, 2$, we verify 
that

\begin{subequations}

\begin{eqnarray}
g_1 (0) &=& g_1 (n T_{\Gamma}) ,  \label {3.6a} \\
&&  \nonumber \\
g_2 (0) &=& g_2 (n T_{\Gamma}) ,  \label {3.6b}
\end{eqnarray}

\end{subequations}

\noindent where $n= 0, 1, 2, \cdots$

\vspace{0.8cm}


\subsection{Cyclic evolution at $\tau = T$}  \label{S3.1}

The vector state (\ref{2.16}) at $\tau = T$ is

\begin{eqnarray}  \label{3.7}
| \psi (T) \rangle = - e^{- \frac{i}{2} (\varepsilon_1 + \varepsilon_2) T}
\; \left[ g_1 (T) | \phi_1; 0\rangle +  g_2 (T) | \phi_2; 0\rangle \right] ,
\end{eqnarray}

\noindent where 

\begin{subequations}

\begin{eqnarray}
g_1 (T) = c_1 (0) \cos(\Gamma \, T) + i \frac{\sin(\Gamma \, T)}{\Gamma} 
\left[ ( \tilde{E}_2 - \frac{\omega}{2} \cos(\theta)) \, c_1(0) 
- \frac{\omega}{2} \sin(\theta) \, c_2 (0) \right] ,
                         \label{3.8a}\\
&&  \nonumber \\
g_2 (T) = c_2 (0) \cos(\Gamma \, T) - i \frac{\sin(\Gamma \, T)}{\Gamma} 
\left[ ( \tilde{E}_2 - \frac{\omega}{2} \cos(\theta)) \, c_2(0) 
+ \frac{\omega}{2} \sin(\theta) \, c_1 (0) \right] .
                         \label{3.8b}
\end{eqnarray}

\end{subequations}

We verify from eqs. (\ref{3.7})-(\ref{3.8b}) that for 
the vector state $|\psi (T)\rangle$ to differ by a 
global phase from $|\psi (0)\rangle$ (see eq.(\ref{3.1})), 
the ratio $\left(\frac{\Gamma}{\omega}\right)$ has to have
particular values or we have to choose special values for
the initial coefficients $c_1 (0)$ and $c_2(0)$.

\vspace{0.3cm}

We begin the discussion of $\tau = T$  with two special cases\cite{ni,zhu}:

\vspace{0.3cm}

\noindent {\bf 1)} $\Gamma = n \omega$, \hspace{0.5cm} $n = 1, 2 , 3, \cdots$

For any initial vector state (\ref{2.14}), where $c_1 (0)$
and $c_2 (0) \in \mathbb{C}$, the vector state at $t=T$ is written
as  eq.(\ref{3.1})  with

\begin{eqnarray}  \label{3.1.1}
\phi = - \pi - \frac{1}{2} (\varepsilon_1 + \varepsilon_2) T.
\end{eqnarray}

To derive the Aharonov- Anandan phase  defined 
in eq.(\ref{3.3a}), we use eq.(\ref{A.3}) to calculate the 
integral of the expectation value of the energy in the state
$|\psi (t)\rangle$ during the interval $t \in [0, T]$. 
In the general case, the coefficient $c_1 (0) \in \mathbb{C}$
and we write it as  $c_1 (0) = |c_1 (0)| e^{i \delta_1}$,
$\delta_1 \in \mathbb{R}$.
 
Subtracting the dynamical phase from the global phase
(\ref{3.1.1}), we obtain

\begin{eqnarray}  \label{3.1.2}
\beta &=& - \pi - \frac{\pi}{\cos(\theta)} \left( \frac{\Delta \varepsilon}{\omega} \right)
\left[ (|c_2 (0)|^2 - |c_1 (0)|^2) \left( 1- \frac{\sin^2(\theta)}{4n^2} \right) \right.
   \nonumber \\
&& \hspace{1cm}
\left. + \frac{\sin(\theta)}{2n} \; \sqrt{ 1 - \frac{\sin^2(\theta)}{4n^2}} \; \;
|c_1 (0)| \;\; [ c_2(0) e^{-i \delta_1} +   c_2^*(0) e^{i \delta_1} ]
\right]  .
\end{eqnarray}

In the adiabatic limit ($\Delta \varepsilon \gg \omega$), with $c_2 (0) =0$,
we recover from eq.(\ref{3.1.2}) the Berry's phase acquired by the 
instantaneous eigenstate of energy $|\phi_1; t\rangle$ (see eq.(26)
of Ref.\cite{ajp2000}). In the same limit, the Aharonov-Anandan
phase with $c_1(0) =0$ is equal to Berry's phase associated to 
the state $|\phi_2; t\rangle$ (see eq.(28) of Ref.\cite{ajp2000}). In each
case, the Berry's phase, or equivalently the Aharonov-Anandan phase in
the adiabatic limit, is independent of the initial coefficients 
$c_1 (0)$ and $c_2(0)$.

The $\beta$ phase (\ref{3.1.2}) is acquired by any vector state 
$|\psi (t)\rangle$ at $\tau = T$ when $\Gamma = n \omega$, 
$n = 1, 2, \cdots$, but its value depends on the the initial
coefficients $c_1 (0)$ and $c_2 (0)$, differently from what
is stated in Refs.\cite{imai1} and \cite{imai2}.

In eq.(\ref{3.1.2}) we write the $\beta$ phase as a function of
$\cos(\theta)$ and $\sin(\theta)$ in order to show that 
in the quasi-resonant regime 
$(\frac{\Delta \varepsilon}{\omega} \approx 1)$ this 
phase is not equal to the solid angle enclosed  by the 
evolution of the effective  magnetic field 
(\ref{2.6a})-(\ref{2.6c}).

From the expression of the Rabi frequency (\ref{2.19})
we  obtain the value of the coupling $D_0$ between the
matter and the electric field such that   the  condition 
$\Gamma = n \omega$, $n = 1, 2, \cdots$ is satisfied

\begin{eqnarray}  \label{3.1.3}
D_0^2 = n^2 \omega^2 - \frac{1}{4} (\Delta \varepsilon + \omega)^2 ,
\hspace{0.5cm} n= 1, 2, \cdots
\end{eqnarray}

At this point, we restrict the discussion  to the quasi-resonant 
regime ($\omega = \Delta \varepsilon + \delta, \frac{\delta}{\Delta \varepsilon} \ll 1$).
In this regime, eq.(\ref{3.1.3}) becomes, up to first order in
$(\frac{\delta}{\Delta \varepsilon})$, 

\begin{eqnarray}  \label{3.1.4}
 \left( \frac{D_0}{\Delta \varepsilon}\right)^2 \approx 
 (n^2 -1) + (2n^2 -1) \; \left(\frac{\delta}{\Delta \varepsilon}\right) , 
 \hspace{0.5cm} n = 1, 2, \cdots
\end{eqnarray}

For $n=1$, eq.(\ref{3.1.4}) is 

\begin{eqnarray}  \label{3.1.4a}
 \left( \frac{D_0}{\Delta \varepsilon}\right)^2 \approx \frac{\delta}{\Delta \varepsilon} .
 \hspace{0.5cm} n = 1, 2, \cdots
\end{eqnarray}

\noindent Since $\Delta \varepsilon >0$, the previous equation
only has solution for $\delta >0$. The condition (\ref{3.1.4a})
is satisfied by the limit of weak coupling between  matter and
electric field $(D_0 \ll \frac{\Delta \varepsilon}{2})$. On the 
other hand, for $n\ge 2$, the equality (\ref{3.1.4}) can not be 
accomplished by a weak coupling.

\vspace{0.3cm}

\noindent {\bf 2)} $\Gamma = m \omega$, \hspace{0.5cm} 
$m = \frac{1}{2},  \frac{3}{2}, \frac{5}{2}, \cdots$

In this case, any vector state (\ref{2.14}), with $c_1 (0)$ and 
$c_2 (0) \in \mathbb{C}$, acquires at $t= T$ the global phase

\begin{eqnarray}  \label{3.1.5}
\phi = - \frac{1}{2} \; (\varepsilon_1 + \varepsilon_2) T .
\end{eqnarray}

Subtracting the contribution of the dynamical phase from the 
global phase (\ref{3.1.5}), we obtain the Aharonov-Anandan
phase 

\begin{eqnarray}  \label{3.1.6}
\beta &=& - \frac{\pi}{\cos(\theta)} \; \left(\frac{\Delta \varepsilon}{\omega}\right) \; \;
\left[ (|c_2 (0)|^2 - |c_1 (0)|^2) \; \left(1 - \frac{\sin^2 (\theta)}{4 m^2}\right)    \right.   \nonumber \\
&& \hspace{1cm} \left.
+ \frac{\sin(\theta)}{2m}  \; \; \sqrt{1 - \frac{\sin^2 (\theta)}{4 m^2}}
\;\; |c_1 (0)| \; ( c_2(0) e^{-i \delta_1} +c_2^*(0) e^{i \delta_1} ) \right],
\end{eqnarray}

\noindent with $m = \frac{1}{2},  \frac{3}{2}, \frac{5}{2}, \cdots$

Again we verify that the $\beta$ phase depends on the initial vector
state. We remind that $c_1 (0) = |c_1 (0)| e^{i \delta_1}$, with
$\delta_1 \in \mathbb{R}$. Comparing the phases (\ref{3.1.2}) and (\ref{3.1.6}), 
we see that they  have similar expressions up to a $\pi$-phase.

To satisfy the condition $\Gamma = m \omega$, 
$m = \frac{1}{2},  \frac{3}{2}, \frac{5}{2}, \cdots$, in the
quasi-resonant regime, the coupling $D_0$ has to be

\begin{eqnarray}  \label{3.1.7}
\left( \frac{D_0}{\Delta \varepsilon}\right)^2 \approx 
 (m^2 -1) + (2m^2 -1) \; \left(\frac{\delta}{\Delta \varepsilon}\right) , 
\hspace{0.5cm} m = \frac{1}{2},  \frac{3}{2}, \frac{5}{2}, \cdots
\end{eqnarray}

\noindent We verify that eq.(\ref{3.1.7}) does not have solution
for $m= \frac{1}{2}$. For $m = \frac{3}{2}, \frac{5}{2}, \cdots$,
we see again that the condition (\ref{3.1.7}) can not be fulfilled 
by a weak  coupling between the matter and the electric field. 

\vspace{0.7cm}

We now consider a third condition among the frequencies $\Gamma$
 and $\omega$, which were not previously discussed in Refs.\cite{ni,zhu}.
 
\vspace{0.3cm}

\noindent {\bf 3)} $\Gamma \not= l \omega$, \hspace{0.5cm} $l = \frac{1}{2}, 1, \frac{3}{2}, 2 \cdots$

For the vector state to satisfy the cyclic condition (\ref{3.1})  at $\tau = T$,
the initial coefficients $c_1 (0)$ and $c_2 (0)$ has to satisfy the coupled
linear equations,

\begin{eqnarray}  \label{3.1.8}
{\bf M} \; \left(\begin{array}{c}
                   c_1 (0) \\
                   c_2(0)
                  \end{array}\right)  = 
        \left( \begin{array}{c}
                   0 \\
                   0
                  \end{array}  
            \right)  .
\end{eqnarray}

The entries of the matrix {\bf M} are,

\begin{subequations}

\begin{eqnarray}  \label{3.1.9}
M_{11} &=& \Gamma \tilde{E}_2 (\cos(\Gamma T)  + e^{i\gamma})  
+ i \sin(\Gamma T) \; \left(\tilde{E}_2^2 + \frac{\omega \Delta \varepsilon}{4} \right) ,
                  \label{3.1.9a} \\
M_{12} &=&  M_{21} =  - i \, \frac{\omega D_0}{2} \sin(\Gamma T)  \label{3.1.9b} , \\
M_{22} &=& \Gamma \tilde{E}_2 (\cos(\Gamma T)  + e^{i\gamma})  
- i \sin(\Gamma T) \; \left(\tilde{E}_2^2 + \frac{\omega \Delta \varepsilon}{4} \right) .
                  \label{3.1.9c}
\end{eqnarray}

\end{subequations}

The phase $\gamma$ is defined as

\begin{eqnarray}  \label{3.1.10}
\gamma \equiv \phi + \frac{(\varepsilon_1 + \varepsilon_2)}{2} \, T.
\end{eqnarray}

Eq.(\ref{3.1.8}) has non-trivial solution only if
that

\begin{eqnarray} \label{3.1.11}
det({\bf M}) = 0.
\end{eqnarray}

From the entries of matrix {\bf M}, we obtain

\begin{eqnarray} \label{3.1.12}
det({\bf M}) = \Gamma^2 \, \tilde{E}_2^2 \; 
\left(1 + 2 \cos(\Gamma T) \, e^{i \gamma} + e^{2 i \gamma} \right) .
\end{eqnarray}

The r.h.s. of the eq.(\ref{3.1.12}) satisfies condition
(\ref{3.1.11}) when

\begin{eqnarray} \label{3.1.13}
\gamma = \pi \pm \Gamma T ,
\end{eqnarray}

\noindent and consequently we obtain the global phase

\begin{eqnarray}  \label{3.1.14}
\phi = \pi \pm - \frac{(\varepsilon_1 + \varepsilon_2) T}{2} .
\end{eqnarray}

Solving eq.(\ref{3.1.8})  at $\phi = \pi \pm - \frac{(\varepsilon_1 + \varepsilon_2) T}{2} $
and imposing that the coefficients $c_1 (0)$ and $c_2 (0)$ satisfy the 
normalization condition, we obtain

\begin{subequations}

\begin{eqnarray}
c_1 (0) &=& \frac{\omega D_0 e^{i \delta_1}} {\left[\omega^2 D_0^2 
+ 4 \; [ \tilde{E}_2 \; (\tilde{E}_2 \mp \Gamma) 
+ \frac{\omega \Delta \varepsilon}{4}]^2 \right]^{\frac{1}{2}}} ,
     \label{3.1.15a}  \\
  \nonumber  \\
c_2 (0) &=& \frac{ 2 [ \tilde{E}_2 \; (\tilde{E}_2 \mp \Gamma) 
+  \frac{\omega \Delta \varepsilon}{4}] \; e^{i \delta_1}}   
{\left[\omega^2 D_0^2 
+ 4 \; [ \tilde{E}_2 \; (\tilde{E}_2 \mp \Gamma) 
+ \frac{\omega \Delta \varepsilon}{4}]^2 \right]^{\frac{1}{2}}} ,
     \label{3.1.15b}
\end{eqnarray}

\end{subequations}

\noindent with $\delta_1 \in \mathbb{R}$. From the previous equations, 
we verify that only for particular values of the coefficients $c_1 (0)$
and $c_2 (0)$ the vector state satisfy the condition (\ref{3.1})
at $\tau = T$ when 
$\Gamma \not= l \omega$, $l = \frac{1}{2}, 1, \frac{3}{2}, 2 \cdots$

The Aharonov-Anandan phase acquired by the vector state at $\tau = T$ when 
$\phi = - \frac{(\varepsilon_1 + \varepsilon_2) T}{2} +\pi \pm \Gamma T$ is

\begin{eqnarray} \label{3.1.16}
\beta &=& \pi \pm \Gamma T - \frac{\pi}{\cos(\theta)} \left(\frac{\Delta \varepsilon}{\omega} \right)
\; \left\{(|c_2 (0)|^2 - |c_1 (0)|^2) \;  \left[ 1-  \frac{\omega^2}{4 \Gamma^2}
\; \sin^2 (\theta)  \; \left(1 - \frac{\omega}{4\pi \Gamma} \sin(2 \Gamma T)\right) \right]
\right.    \nonumber \\
&+& \left. 
 \frac{\omega}{\Gamma} \left(\frac{1}{2} - \frac{\omega}{8\pi \Gamma}\, \sin(2\Gamma T)\right)
 \sin(\theta) \; \sqrt{1-  \frac{\omega^2}{4 \Gamma^2}\; \sin^2 (\theta)} \; \;
 |c_1 (0)| \, (c_2 (0) e^{-i\delta_1} + c_2^* (0) e^{i\delta_1}) \right\}.
        \nonumber\\
\end{eqnarray}

\noindent The coefficients $c_1 (0)$ and $c_2 (0)$ in the previous equation are
given by expression (\ref{3.1.15a}) and (\ref{3.1.15b}) respectively. In this case,
the $\beta$ phase is independent of the value of $\delta_1 \in \mathbb{R}$.

\vspace{0.5cm}

As a final comment on the cyclic evolution of the vector state 
during the period  of the external electric field, we should 
notice that if the ratio

\begin{eqnarray} \label{3.1.17}
\frac{\Gamma}{\omega} = \frac{m}{2n}, \hspace{0.5cm} m, n = 1, 2, 3, \cdots
\end{eqnarray}

\noindent then at $\tau = n T$, the vector state $|\psi (nT)\rangle$ 
satisfies the condition (\ref{3.1}) for any initial vector state,
and the global $\phi$ phase is

\begin{eqnarray}  \label{3.1.18}
\phi = - \frac{1}{2} (\varepsilon_1 + \varepsilon_2) nT + (m - n) T,
\end{eqnarray}

\noindent with $m, n = 1, 2, 3, \cdots$.

\vspace{0.8cm}


\subsection{Cyclic evolution at  $\tau \not= n T$, $n = 1, 2, 3, \cdots$}  \label{S3.2}

When the cyclic condition (\ref{3.1}) is verified at $\tau \not= n T$,
$n = \frac{1}{2}, 1, \frac{3}{2}, 2, \cdots$, we have 
$|\phi_j; \tau\rangle \not= |\phi_j; 0\rangle$, $ j= 1,2$.
Moreover, we note that for the $\tau \not= n T$, $n = 1, 2, 3, \cdots$, 
the classical electric field does not travel a closed path in its 
parameter space when the initial state vector acquires a phase. 

In order the discuss the condition (\ref{3.1}) at $\tau \not= n T$, it is better 
to write the states $|\psi (0)\rangle$ and $|\psi (\tau)\rangle$ in the basis of 
the eigenstates of hamiltonian ${\bf H}_0$. To fulfill the condition (\ref{3.1}), 
the initial coefficients  $c_1 (0)$ and $c_2 (0)$ in the initial vector 
state (\ref{2.14}) have to satisfy the coupled linear equations:

\begin{eqnarray}  \label{3.2.1}
\tilde{\bf M} \; \left(\begin{array}{c}
                   c_1 (0) \\
                   c_2(0)
                  \end{array}\right)  = 
        \left( \begin{array}{c}
                   0 \\
                   0
                  \end{array}
            \right)  .
\end{eqnarray}

The entries of the matrix $\tilde{\bf M}$ are

\begin{subequations}

\begin{eqnarray}
\tilde{M}_{11} &=& \sin(\frac{\theta}{2}) \, [e^{i \tilde{\gamma}} - \cos(\Gamma \tau)] \; 
+ i \frac{\sin(\Gamma  \tau)}{\Gamma} \left[ - \sin(\frac{\theta}{2}) (\tilde{E}_2 
              - \frac{\omega \cos(\theta)}{2})
- \frac{\omega \sin(\theta)}{2} \cos(\frac{\theta}{2})   \right] ,  \nonumber  \\
&&  \nonumber  \\
&&  \label {3.2.2a} \\
\tilde{M}_{12} &=& - \cos(\frac{\theta}{2}) \, [e^{i \tilde{\gamma}} - \cos(\Gamma \tau)] \; 
-  i \frac{\sin(\Gamma  \tau)}{\Gamma} \left[\cos(\frac{\theta}{2}) (\tilde{E}_2 
              - \frac{\omega \cos(\theta)}{2})
- \frac{\omega \sin(\theta)}{2} \sin(\frac{\theta}{2})   \right] , \nonumber  \\
&&  \label {3.2.2b} \\
&&  \nonumber  \\
\tilde{M}_{21} &=& \cos(\frac{\theta}{2}) \, [- e^{i (\tilde{\gamma} - \omega \tau)} + \cos(\Gamma \tau)] \; 
+ i \frac{\sin(\Gamma  \tau)}{\Gamma} \left[\cos(\frac{\theta}{2}) (\tilde{E}_2 
              - \frac{\omega \cos(\theta)}{2})
- \frac{\omega \sin(\theta)}{2} \sin(\frac{\theta}{2})   \right] , \nonumber  \\
&&  \label {3.2.2c} \\
&&  \nonumber  \\
\tilde{M}_{22} &=& \sin(\frac{\theta}{2}) \, [- e^{i (\tilde{\gamma} - \omega \tau)} + \cos(\Gamma \tau)] \; 
+ i \frac{\sin(\Gamma  \tau)}{\Gamma} \left[ - \sin(\frac{\theta}{2}) (\tilde{E}_2 
              - \frac{\omega \cos(\theta)}{2})
- \frac{\omega \sin(\theta)}{2} \cos(\frac{\theta}{2})   \right] . \nonumber  \\
&&  \label {3.2.2d} 
\end{eqnarray}

\end{subequations}

\noindent The phase $\tilde{\gamma}$ is defined as

\begin{eqnarray}  \label{3.2.3}
\tilde{\gamma} (\tau) \equiv \phi + \frac{(\varepsilon_1 + \varepsilon_2 + \omega) \tau}{2}.
\end{eqnarray}

The coupled eqs.(\ref{3.2.1}) have non-trivial solution only if
$det (\tilde{\bf M}) =0$. By direct calculation we obtain

\begin{eqnarray}  \label{3.2.4}
det(\tilde{\bf M}) = -1 + [ \cos(\Gamma \tau) \; ( e^{i \tilde{\gamma}} + e^{i (\tilde{\gamma} - \omega \tau)} )
- e^{i (2\tilde{\gamma} - \omega \tau)}] 
-  i \frac{\sin(\Gamma \tau)}{2 \Gamma} \, (\omega + \Delta \varepsilon) \,
               [e^{i \tilde{\gamma}} - e^{i (\tilde{\gamma} - \omega \tau)} ] . \nonumber\\
&& 
\end{eqnarray}

All the cases discussed in subsection \ref{S3.1} satisfy identically
the condition $det (\tilde{\bf M}) = 0$ when the relations between $\Gamma$ 
and $\omega$ are substituted on the r.h.s. of eq.(\ref{3.2.4}).

\vspace{0.5cm}

We will not discuss the general case when  (\ref{3.2.4})  is null
for the arbitrary  value of $\tau$.

Up to the end of this subsection, we take $\tau = T_{\Gamma}$,
$T_{\Gamma}$ being the period associated to the Rabi's frequency.

The vector state at $\tau = T_{\Gamma}$ is

\begin{eqnarray}   \label{3.2.5}
|\psi (T_{\Gamma}) \rangle &=& e^{- \frac{i}{2} (\varepsilon_1 + \varepsilon_2 + \omega) T_{\Gamma}}
\left\{ \left[ - \sin(\frac{\theta}{2}) \, c_1 (0) + \cos(\frac{\theta}{2}) \, c_2 (0)\right]  | 1 \rangle
   \right.     \nonumber \\
&& \hspace{0.6cm}  \left.
+ e^{- i \phi_0} \, e^{ i \omega T_{\Gamma}} \;  
 \left[\cos(\frac{\theta}{2}) \, c_1 (0)  +  \sin(\frac{\theta}{2}) \, c_2 (0) \right] \, | 2 \rangle
\right\} .
\end{eqnarray}

The vector state (\ref{3.2.5}) is equal to the initial state (\ref{2.14}) up to 
a global phase, independently of the values of $c_1 (0)$ and $c_2 (0)$, if

\begin{eqnarray}  \label{3.2.6}
\Gamma = \frac{\omega}{n}, \hspace{0.5cm}  n = 1, 2, 3, \cdots
\end{eqnarray}

\noindent The previous relation is identical to $T_{\Gamma} = n T$.

In the  quasi-resonant regime, eq.(\ref{3.2.6}) implies 

\begin{eqnarray}  \label{3.2.7}
\left(\frac{D_0}{\Delta \varepsilon}\right)^2  \approx  \left( \frac{1}{n^2} -1\right)
+ \left(\frac{2}{n^2} -1\right) \; \left(\frac{\delta}{\Delta\varepsilon}\right),
   \hspace{0.5cm} n= 1, 2,3, \cdots
\end{eqnarray}

\noindent Eq.(\ref{3.2.7}) has  solution  only for $n=1$ when 
$\delta >0$.

The global phase $\phi$ acquired by any vector state at 
$\tau = T_{\Gamma}$, for $n=1$ in the condition (\ref{3.2.6}) is  

\begin{eqnarray}  \label{3.2.8}
\phi = - \frac{1}{2} \, (\varepsilon_1 + \varepsilon_2) T_{\Gamma} - \pi.
\end{eqnarray}

The Aharonov-Anandan phase in this case is

\begin{eqnarray}  \label{3.2.9}
\beta &=& -  \pi - \frac{\pi}{\cos(\theta)} \, \left( \frac{\Delta \varepsilon}{\omega} \right) \;
\left[ (|c_2(0)|^2 - |c_1(0)|^2) \; \left( 1 - \frac{\sin^2 (\theta)}{4} \right)
    \right.  \nonumber \\
&& \hspace{0.8cm}  \left.  
+ \frac{\sin(\theta)}{2}  \; \sqrt{1 - \frac{\sin^2 (\theta)}{4}} \;
|c_1 (0)| \, (c_2 (0) e^{-i \delta_1} + c_2^* (0) e^{i \delta_1}) \right] .
\end{eqnarray}

\noindent  This $\beta$ phase also depends on the initial vector state.

\vspace{0.5cm}

As our final case we continue to have $\tau = T_{\Gamma}$
but $T_{\Gamma} \not= n T$, with $n= 1, 2, 3, \cdots$

The vector state $|\psi (T_{\Gamma})\rangle$ continues to be 
given by expression (\ref{3.2.5}). Since $T_{\Gamma} \not= n T$,
$ n =1, 2, \cdots$,  we verify  that is not possible for 
any coefficients $c_1 (0)$ and $c_2 (0) \in \mathbb{C}$  
that $|\psi (T_{\Gamma}\rangle$ satisfy the condition 
(\ref{3.1}).

At $\tau = T_{\Gamma}$, we decompose the $det(\tilde{M})$ in 
its real and imaginary parts,

\begin{eqnarray}  \label{3.2.10}
det(\tilde{M}) &=& 2 \sin(\frac{\gamma^{\prime}}{2}) \;  \left\{
\left[-  \sin(\frac{\gamma^{\prime}}{2}) +  \sin(\frac{3\gamma^{\prime}}{2} - \omega T_{\Gamma}) \right]
     \right.    \nonumber   \\
&& \hspace{1.5cm}   \left.
 + 2 i \, \sin(\frac{\gamma^{\prime} - \omega T_{\Gamma}}{2})
   \, \sin(\gamma^{\prime} - \frac{\omega T_{\Gamma}}{2}) 
        \right\}  ,
\end{eqnarray}

\noindent where $\gamma^{\prime} = \tilde{\gamma} (T_{\Gamma})$.

To have $det(\tilde{M}) = 0 $, the real and imaginary parts 
on the r.h.s. of (\ref{3.2.10}) have to be simultaneously
null. For arbitrary value of $\omega T_{\Gamma}$, one
solution to $det(\tilde{M}) = 0 $ is

\begin{eqnarray} \label{3.2.11}
\gamma^{\prime} \sim 0, \hspace{1cm} \mbox{mod} (2\pi) .
\end{eqnarray}

To calculate the values of the initial coefficients $c_1(0)$ 
and $c_2(0)$ such that the vector $|\psi (T_{\Gamma})\rangle$ 
acquires the global phase

\begin{eqnarray}  \label{3.2.12}
\phi = - \frac{(\varepsilon_1 + \varepsilon_2) T_{\Gamma}}{2}
- \frac{\omega T_{\Gamma}}{2}
\end{eqnarray}

\noindent and the Aharonov-Anandan phase

\begin{eqnarray}  \label{3.2.13}
\beta &=& - \frac{\pi \omega}{\Gamma} - \frac{\pi}{\cos(\theta)} \, 
\left(\frac{\omega}{\Gamma}\right)\; \left(\frac{\Delta \varepsilon}{\omega}\right)
\; \left\{ [|c_2(0)|^2 - |c_1(0)|^2] \; \left(1 - \frac{\omega^2}{4 \Gamma^2} \sin^2 (\theta) \right)   \right.
                    \nonumber   \\
&+&  \left.
\frac{\omega}{2 \Gamma} \sin(\theta) \; \sqrt{1 - \frac{\omega^2}{4 \Gamma^2} \sin^2 (\theta)}
\;\;  |c_1(0)| \, [c_2 (0) e^{- i \delta_1} + c_2^* (0) e^{i \delta_1}] 
      \right\} ,
\end{eqnarray}

\noindent the solution (\ref{3.2.11}) has to be replaced  on the
coupled eq.(\ref{3.2.1}).

Besides the solution (\ref{3.2.11}), the real part of the r.h.s. 
of (\ref{3.2.10}) has the common root:

\begin{eqnarray}  \label{3.2.14a}
\gamma^{\prime} \sim  \omega T_{\Gamma}, \hspace{1cm} \mbox{mod} (2 \pi) ,
\end{eqnarray}

\noindent valid for arbitrary value of $\omega T_{\Gamma}$.

The vector state at $\tau = T_{\Gamma}$, for particular 
values of $c_1(0)$ and $c_2 (0)$, acquires the global phase

\begin{eqnarray}  \label{3.2.17}
\phi \sim -  \frac{1}{2} (\varepsilon_1 + \varepsilon_2) T_{\Gamma}
 + \frac{\omega T_{\Gamma}}{2} 
   , \hspace{1cm} \mbox{mod} (2 \pi)  .
\end{eqnarray}

The Aharonov-Anandan phase (the $\beta$ phase)  in this case is equal to

\begin{eqnarray}  \label{3.2.18}
\beta &=&  \frac{\pi \omega}{\Gamma} - \frac{\pi}{\cos(\theta)} \, 
\left(\frac{\omega}{\Gamma}\right)\; \left(\frac{\Delta \varepsilon}{\omega}\right)
\; \left\{ [|c_2(0)|^2 - |c_1(0)|^2] \; \left(1 - \frac{\omega^2}{4 \Gamma^2} \sin^2 (\theta) \right)   \right.
                    \nonumber   \\
&+&  \left.
\frac{\omega}{2 \Gamma} \sin(\theta) \; \sqrt{1 - \frac{\omega^2}{4 \Gamma^2} \sin^2 (\theta)}
\;\;  |c_1(0)| \, [c_2 (0) e^{- i \delta_1} + c_2^* (0) e^{i \delta_1}] 
      \right\}  .
\end{eqnarray}


\section{ Conclusions } \label{S4}

We study the dynamical evolution of a two-level model coupled
to a classical linearly polarized monochromatic electric field in the
RWA. We map this model onto a spin-1/2 model coupled to an
effective magnetic field. The dynamics of the latter model
has already been studied in the literature\cite{ajp2000,gonzalo},
for any regime. In particular, its Aharonov-Anandan phases were studied in 
Refs.\cite{ni,zhu} to the condition $\Gamma = n \omega$, 
$n = \frac{1}{2}, 1, \frac{3}{2}, 2, \cdots$

Our results on the dynamics of the RWA of the electric two-level
model (see hamiltonian (\ref{2.2})) are valid for any value of the
ratio $\frac{\omega}{\Delta \varepsilon}$. In particular, this model in 
the quasi-resonant regime is a good description
of the coupling between matter and an electric field in the electric
dipole approximation, being a candidate for modeling a
gate in quantum computation\cite{nielsen,xin}.

We show that when  a cycle of the vector state
occurs within a cycle of the external electric field, the
Aharonov-Anandan phase depends on the initial vector state, unless
the quantum system evolves adiabatically. Our results are in
disagreement with Refs.\cite{imai1,imai2}.

We also show that the periodic behavior of the vector state (see
eq.(\ref{3.1.1})) can also happen within the
period $T_{\Gamma}$ associated to Rabi's frequency (\ref{2.19}).
Again, the Aharonov-Anandan phase depends on the initial 
coefficients of the initial vector state.

In both cases, we verified that in the non-adiabatic regime the
Aharonov-Anandan phase depends on the initial state and on the
interval of time for the vector state to return to its initial
state, up to a global phase. Although our results are valid for 
any value of the
$\frac{\omega}{\Delta \varepsilon}$, part of our results are 
discussed in the quasi-resonant regime, where we expect our results
to be applicable to the modeling of a quantum computation device. 

Our results are easily reinterpreted in terms of the spin-1/2 model  
in the presence of a rotating magnetic field. They extend 
previous discussions\cite{bulgac,ni,zhu} on the Aharonov-Anandan 
phases of this model.

\vspace{1cm}

The authors thank E.V. Corr\^ea Silva for the careful reading
of part of this manuscript. M.T.T. is in debt with S.A. Dias for 
the nice discussions on Ref.\cite{aharonov}.
M.T. Thomaz  (Fellowship CNPq, Brazil, Proc.No.: 30.0549/83-FA) 
thanks CNPq  for partial financial support.




\setcounter         {equation}{0}
\def\theequation{A.\arabic{equation}}
\def\thesection{Appendix A}

\section{The average energy in any time-dependent vector 
                     state} \label{Appendix_A}

Ref.\cite{aharonov} writes the periodic condition of the vector state 
$| \psi (t)\rangle$ at $t = \tau$  as 

\begin{subequations}

\begin{eqnarray}
| \psi (\tau) \rangle &=& e^{i \phi} \, | \psi(0) \rangle  \label{A.1a} \\
                       &=&  e^{i \beta} \;
 e^{-i \int_0^{\tau} \, \langle \psi (t)| {\bf H}_e (t) | \psi (t) \rangle \, dt} \;\,
        | \psi (0) \rangle ,  \label{A.1b}
\end{eqnarray}

\end{subequations}

\noindent where the dynamical phase is diminished from the global phase
$\phi$.

\vspace{0.3cm}

We assume the initial vector state (\ref{2.14}) and let $c_1 (0)$ and 
$c_2 (0) \in \mathbb{C}$. The coefficient $c_1 (0)$ is written as

\begin{eqnarray}   \label{A.2}
c_1 (0) = | c_1 (0)| \, e^{i \delta_1} ,
\end{eqnarray}

\noindent with $\delta_1 \in \mathbb{R}$.

The expectation value of the energy in the vector state (\ref{2.16}),
at any instant $t$, is

\begin{eqnarray}  \label{A.3}
\langle \psi (t)| {\bf H}_e (t) | \psi (t) \rangle &=&  \frac{(\varepsilon_1 + \varepsilon_2)}{2}
+ \tilde{E}_2  \left\{  [ \, |c_2 (0)|^2 - |c_1 (0)|^2 ]   \right. \nonumber \\
&+& \frac{i}{2} \frac{\omega}{\Gamma} \sin(\theta) \, |c_1 (0)| \,  \sin(2\Gamma t) 
   [ c_2(0) e^{- i \delta_1} - c_2^* (0) e^{ i \delta_1}]   \nonumber \\
&& \hspace{-2cm} 
+ \sin^2 (\Gamma t) \left[ - \frac{\omega^2}{ 2 \Gamma^2} [ |c_2 (0)|^2 - |c_1 (0)|^2 ] \, \sin^2 (\theta)
      \right.   \nonumber   \\
 &+&  \left. \left.
  \frac{\omega}{\Gamma} \sin (\theta) \sqrt{1 - \frac{\omega^2}{4 \Gamma^2} \sin^2 (\theta)}  \;
  |c_1 (0)| \, [ c_2(0) e^{- i \delta_1} + c_2^* (0) e^{ i \delta_1}]       \right]  \right\}  .
        \nonumber \\
\end{eqnarray}


\end{document}